# Nanospheres with Patches Arranged in Polyhedrons from Self-Assembly of Solution-State Diblock Copolymers under Spherical Confinement

*Jiaping Wu, Xin Wang, Zheng Wang, Yuhua Yin, Run Jiang, Yao Li\*, Baohui Li \**

School of Physics, Key Laboratory of Functional Polymer Materials of Ministry of Education, Nankai University, and Collaborative Innovation Center of Chemical Science and Engineering (Tianjin), Tianjin, 300071, China

**for Table of Contents use only**

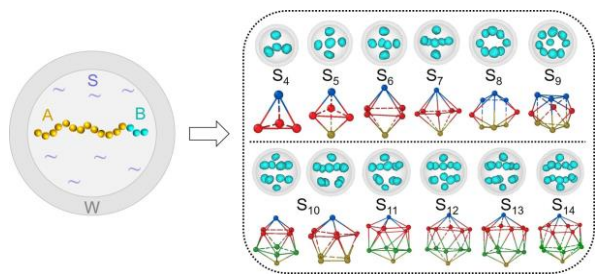

# Abstract

Self-assembly of sphere-forming amphiphilic diblock copolymer solutions confined in spherical nanopores is investigated using simulated annealing technique. Structures with 1–21 solvophobic domains are obtained and phase diagrams are constructed for two types of cases of different pore-surface/copolymer interactions. Various polyhedrons, some of them having high symmetry, such as Platonic solids of regular tetrahedron, octahedron and icosahedron and eight Johnson solids, are identified when connect the centers of the solvophobic domains in structures with various number of solvophobic domains. These polyhedrons are usually with all or most of their faces being in triangular shape, which may relieve the chain stretching. In one of the two types of cases, all structures with 1–14 outermost solvophobic domains have a stable region in the phase diagram, whereas in the other type of cases structures with some number of solvophobic domains are not observed as stable due to the relatively weak constraint of the pore surface to the copolymer. Furthermore, at a given pore size, the number of the solvophobic domains changes nonmonotonically or keeps unchanged with increasing the strength of the pore-surface/copolymer interaction for the former and the latter type of cases, respectively. It is found that the solvophilic blocks present a larger stretching when all of the solvophobic domains are attached to the pore surface or when the polymer concentration is low. Our results may provide fundamental understanding about the relationship between the confinement conditions, solvent conditions and the self-assembled structures.

# 1 Introduction

Block copolymers can self-assemble into phase-separated periodic nanostructures, such as lamellae, gyroid, cylinders, spheres, etc., and have attracted much scientific interest due to the potential application of these structures in various area[1–4], such as in nanolithography[3] and material science[4]. Confinement can be used to break the symmetry of structures formed in the self-assembly of block copolymers and hence generate novel structures that cannot be obtained in bulk[3,5,6]. In general, confinement can be classified into one-dimensional (1D) planar confinement[3,7–10], two-dimensional (2D) cylindrical confinement[5,11–16] and three-dimensional (3D) spherical confinement[6,17–21] based on the confining geometries. Previous extensive studies showed that frustration induced by 1D confinement of block copolymers in thin films can lead to different orientations and adjustable periods of structures depending on the surface-block interactions and film thickness[22,23]. While 2D cylindrical confinement and 3D spherical confinement could induce frustration and distortions that cannot be observed in bulk or 1D confinement, due to the curvature of the confining pore in the 2D and 3D confinement. And a large number of novel and frustrated structures were found[5,9,20,24–27], such as cylindrical (spherical) - concentric lamellar structures and stacked lamellae, as well as non-lamellar structures[20,27] including Janus-type, tennis-ball-, mushroom-, wheel-, or screwlike structures[20] for the bulk lamella-forming diblock copolymers, while for the bulk cylinder-forming diblock copolymers, single-helices, double-helices, triple-helices, stacked toroid structures[5,15,28–30], or curved cylinders[9,24,27] were observed in cylindrical or spherical pore confinement, respectively.

Compared to the large number of works of 3D confinement of the lamella- and cylinder-forming diblock copolymers, however, only two studies were reported on the self-assembly of sphere-forming diblock copolymers confined in 3D spherical nanopores. 30 years ago, Reffner experimentally observed irregularly packed spheres when sphere-forming diblock copolymers were confined in droplets and heated to remove solvent[24]. Recently, Zhao et al. studied the phase behavior of both bulk cylinder-forming and sphere-forming diblock copolymers confined in the spherical pores using self-consistent field theory[31]. They constructed a phase diagram as a function of the pore radius and the volume fraction of the copolymer, and found a variety of interesting polyhedral shapes formed by the packing of spherical domains of number being from four to twelve. Furthermore, the number of spherical domains changes discontinuously with increasing the pore size and structures with higher symmetry have a larger stable region in the phase diagram[31].

Compared to the melt system, sphere-forming block copolymer solution should be easier to be obtained by choosing various selective solvents, such as that observed previously[32]. To the best of our knowledge, there are virtually no studies on the self-assembly of sphere-forming block copolymer solutions confined in a spherical nanopore. Previous experimental studies of 2D confinement of block copolymer solutions showed that the dimensions of the nanochannels not only control the packing of spherical micelles but also affect the shape/morphology of individual micelles[33], although the arrangement of spherical micelles in that solution system is similar to that of spherical domains obtained in the block copolymer melt under 2D confinement [15].

In the present work, we study the effect of 3D confinement on the self-assembly of sphere-forming amphiphilic diblock copolymer solutions using simulated annealing technique. The 3D confinement is realized by putting diblock copolymer solutions into a spherical nanopore. Structures with various number of solvophobic domains are obtained and phase diagrams are constructed as a function of the pore radius and the surface-block interaction. Quantitative calculations are performed to understand the details of the obtained structures. Effects of polymer segment concentration on the self-assembled structures and structural details are further investigated. Our simulation results are compared with previous related works.

## 2 Model and Method

The solution-state AB diblock copolymers are modeled by the single-site bond fluctuation model[34–36] and the simulations are performed using simulated annealing method[37,38]. The details of the model and method can be found elsewhere[39]. Asymmetric AB diblock copolymer model molecules of the form $A_{13}B_3$ are used in the study, that is, the degree of polymerization of each A and B block is $N_A=13$ and $N_B=3$. The volume fraction of B-block is $f_B \equiv N_B/N=0.1875$, where $N \equiv N_A+N_B=16$ is the total degree of polymerization of each AB diblock copolymer molecule. In our study, the AB diblock copolymers are dissolved in solution and confined in a spherical nanopore of radius of $R$. The pore is embedded in a simple cubic lattice of volume $V=L^3$ with $L$ being larger than $2R$. The segment concentration is defined as $C_p \equiv nN/V_p$, with $n$ being the number of the model molecules and $V_p$ the total number of lattice sites inside the pore. The model molecules are assumed to be self-avoiding and mutually avoiding in all the simulation processes. The bond length between two consecutive segments in a copolymer molecule is set to be 1 and $\sqrt{2}$, and thus each site has 18 nearest neighbor sites. The starting configuration is generated by putting the model molecules on the lattice sites inside the pore randomly. Only the nearest-neighboring interactions are considered and modeled by assigning an energy $E_{ij}=\varepsilon_{ij}k_B T_{ref}$ to each nearest-neighboring pair of unlike components $i$ and $j$, where $i, j$=A, B, S, W with S being the solvent molecules and W the lattice sites on the pore wall (surface); $\varepsilon_{ij}$ is the reduced interaction energy, $k_B$ is the Boltzmann constant, and $T_{ref}$ is a reference temperature. The A blocks and B blocks are incompatible and the A-B interactions are set to be $\varepsilon_{AB}=1.0$. The A blocks are solvophilic while the B blocks are solvophobic, and the A-S and B-S interactions are set to be $\varepsilon_{AS}=-0.5$ and $\varepsilon_{BS}=0.5$. The solvent-pore surface interactions are set to be zero. For a given $C_p$, two types of cases are studied with the only difference in block-surface interactions: ($\alpha$) $\varepsilon_{AW}=-\varepsilon_{BW}$ and ($\beta$) $\varepsilon_{AW}=0.0$. And $\varepsilon_{BW}$ is varied from 0 to 4.0 with a step of 0.5 for phase diagrams in both types of $\alpha$ and $\beta$ cases. The initial temperature is $T_0=100T_{ref}$, and the annealing schedule and the sampling strategy are the same as those used previously[27]. Under a given set of parameters, 50 to 100 simulations are performed with the only difference being in the random number generator seed. If multiple structures are obtained in our simulations under the given set of parameters, structures with the highest probability of occurrence are considered as stable, while structures with a low probability of occurrence are considered as metastable.

Some quantities are calculated in the study to characterize the details of the morphologies. The nonsphericity of an aggregate or a collection of particles is defined as[40,41] $\kappa \equiv 3(\lambda_x^4+\lambda_y^4+\lambda_z^4)/2(\lambda_x^2+\lambda_y^2+\lambda_z^2)^2 - 1/2$, where $\lambda_x$, $\lambda_y$ and $\lambda_z$ are the three principal moments of the gyration tensor of the aggregate. $\kappa \in [0,1]$, $\kappa=0$ if the aggregate is a perfect sphere and $\kappa=1$ if all the particles in the aggregate are on a line. By mimicking the definition of the index of polydispersity of the polymer molecular weight, the index of polydispersity of the size of each B-domain is defined as $pdi \equiv E(n_B^2)/E^2(n_B)$, where $n_B$ is the number of B-segments inside each B-domain, $E(n_B)$ or $E(n_B^2)$ is the average value of $n_B$ or $n_B^2$ in a system. $pdi=1$ if all the B-domains have a same $n_B$ value while $pdi$ gets larger if the difference between the $n_B$ values of the B-domains gets larger. The normalized mean-square end-to-end distance of chains of length $N$ is calculated as $R^2_e = R^2_{e,1}/R^2_{e,0}$, where $R^2_{e,1}$ and $R^2_{e,0}$ are the mean-square end-to-end distance of the chains and that of the corresponding ideal chain, respectively, and $R^2_{e,0} = b^2(N-1)$ with $b$ being the root-mean-square of bond length ($b=\sqrt{5/3}$ in our model).

## 3 Result and Discussion

In this section, we present our simulation results of structures self-assembled from the sphere-forming amphiphilic diblock copolymer $A_{13}B_3$ solution confined in spherical nanopores, where the A-blocks are solvophilic and the B-blocks are solvophobic, and the reduced interactions used are $\varepsilon_{AB}=1.0$, $\varepsilon_{AS}=-0.5$ and $\varepsilon_{BS}=0.5$. Without confinement, our simulations show that our model system forms spherical structure and spherical micellar morphology, with the solvophobic B-blocks forming spheres or micelle core at segment concentration of $C_p=0.2-0.85$ and $C_p<0.2$, respectively, due to their smaller volume fraction and their solvophobic nature. When $C_p \geq 0.95$, our model system forms cylindrical structure. Under the confinement of spherical pore, structures with various number of solvophobic domains are obtained when $0.1 \leq C_p \leq 0.7$. Snapshots for typical structures are shown (in Figures 1–3) and analyzed.

Two phase diagrams (Figure 4) as a function of the pore radius $R$ and interactions between polymer and pore surface for two types of cases of different polymer-surface interactions are constructed and discussed. Typical quantities including the effective pore radius, nonsphericity and the index of polydispersity of the size of each B-domain, and the normalized mean-square end-to-end distance of chains are calculated and presented (Figures 5 and 6). Results presented in Figures 1–6 are obtained at $C_p$=0.45. Influence of $C_p$ on self-assembled morphologies and typical quantities are shown (in Figure 7) and discussed.

## 3.1 Polyhedral structures

Figure 1 shows the variation of typical snapshots of structures self-assembled from the model amphiphilic diblock copolymer solution confined in spherical nanopores of two types of surfaces with the pore radius $R$, where each structure is named as $S_i$ or $S_{i+1}$ with $i$ (=1, …, 14) being the number of solvophobic B-domains nearest to the pore surface (outermost B-domains) and the "$i+1$" indicating that besides the $i$ outermost B-domains, there is one B-domain located at the pore center. As shown in Figure 1(a, a') for the case of energetically neutral surfaces ($\varepsilon_{AW}=\varepsilon_{BW}=0$), the $i$ outermost B-domains in structure $S_i/S_{i+1}$ are all attached to the pore surface, which is due to the following two reasons. One reason is the entropy-driven attraction of the shorter B-blocks to the neutral surfaces[23,42], and the other is due to the resulting favorable energy when B-domains are attached to the pore surface. This is because the increased contact between B-segments and the pore surface can reduce the unfavorable contact between A- and B-segments and that between B-segments and solvents. Furthermore, each of the outermost B-domain is in a nearly convex-lens shape (not a spherical shape). When the nature of the pore surface changes to attractive to the longer A blocks while repulsive to the shorter B blocks ($-\varepsilon_{AW}=\varepsilon_{BW}=1.0$), all B-domains are of nearly spherical shape and are away from the pore surface as shown in Figure 1(b, b'). A common feature for structures shown in Figure 1(a, a') and Figure 1(b, b') is that the arrangement of the outermost B-domains is the same when the number of the outermost B-domains is the same regardless of whether there is a B-domain located at the pore center or whether the outermost B-domains are attached to the pore surface or not. Therefore, in the following we only describe $S_i$, instead of $S_{i+1}$ for convenience. Notice that for the sake of clarity, the B-domain appearing at the pore center in Figure 1(a') is shown in green color. In Figure 1(c, d, c', d') lines are plotted between the centers of the nearest B-domains to show the arrangements of the outermost B-domains clearly. The center(s) of the B-domain(s) are abbreviated as B-center(s) hereafter. As shown in Figure 1, the B-domain in structure $S_1$ is located at the pore center, the two B-domains in structure $S_2$ are located at the two poles (two sides) of the pore, and the three B-centers in structure $S_3$ constitute an approximate regular triangle. For structures $S_i$ with $4 \leq i \leq 14$, the B-centers constitute polyhedrons with apparent symmetry, as shown in Figure 1(c, d, c', d'). Among these polyhedrons, three of the five Platonic solids (or convex regular polyhedrons) are identified as shown in Figure 1 (labeled with "P"), i.e., regular tetrahedron ($S_4$), regular octahedron ($S_6$) and regular icosahedron ($S_{12}$); and seven Johnson solids[43,44] are identified in the other polyhedron structures, i.e., structure $S_5$ is triangular bipyramid (J12), $S_7$ pentagonal bipyramid (J13), $S_8$ biaugmented triangular prism (J50), $S_9$ triaugmented triangular prism (J51), $S_{10}$ corresponding to two degenerate structures of gyroelongated square bipyramid (J17) and sphenocorona (J86) (to distinguish these two structures, the structure corresponding to J86 is called $S_{10}'$), and $S_{11}$ augmented sphenocorona (J87). Structure $S_{14}$ is the gyroelongated hexagonal bipyramid, and together with the gyroelongated square bipyramid ($S_{10}$) and the gyroelongated pentagonal bipyramid ($S_{12}$) belongs to the gyroelongated bipyramid constructed by elongating an $n$-gonal bipyramid by inserting an $n$-gonal antiprism between its congruent halves.

It is interesting to find that the B-centers are usually distributed into layers. In structures $S_5$, $S_6$ or $S_7$ they are distributed in three layers with the number of B-domains in each layer being [1, $n$, 1] with $n=3$, 4, and 5 in $S_5$, $S_6$ and $S_7$, respectively, as shown in Figure 1c, where the B-centers in these three layers are marked in blue, red and yellow colors, respectively for clarity. While the B-centers in structures $S_{10}$, $S_{11}$, $S_{12}$, $S_{13}$ or $S_{14}$ are distributed in four layers with the number of B-domains in each layer being [1, $n$, $m$, 1] where [$n$, $m$] = [4,4], [5,4], [5,5], [6,5] and [6,6] in structures $S_{10}$, $S_{11}$, $S_{12}$, $S_{13}$ and $S_{14}$, respectively, as shown in Figure 1c', where the B-centers in these four layers are marked in blue, red, green and yellow colors, respectively. Structure $S_{10}'$ can be regarded as formed by deleting one B-center located at the side of the four-B-center-layer in $S_{11}$. Structure $S_9$ can be constructed by attaching an equilateral square pyramid to each of the three square faces of a triangular prism. While structure $S_8$ can be regarded as constructed by randomly deleting an equilateral square pyramid in $S_9$. For structures with the outermost B-domain number being larger than 14, we find that the B-centers still constitute polyhedrons with most of their faces being in triangular shape. These polyhedrons, however, are without apparent symmetry. The anticube (also known as square antiprism) structures having two square faces, have not been observed in our system, not to mention the cubic structures which have six square faces. There are 92 Johnson solids[43] in total, and each of their faces can be in a triangle, tetragon, pentagon, hexagon, octagon, or decagon, and the whole shape of a Johnson solid can be close to or quite deviating from a sphere. While the seven Johnson solids identified in our simulations, are mostly in triangular faces and their whole shape is close to a sphere. Similarly, the three Platonic solids identified in our simulations all have triangular faces. Thus, we conclude that the system prefers structures with triangular faces, since that triangular faces will make the structure being of more faces than that with tetragonal faces or with faces of more sides, and hence the polyhedron constructed by connecting the nearest B-centers is closer to a sphere in shape. On the other hand, the distribution of the B-domains may be more uniform in structures with triangular faces and thus the stretching of the majority block is relatively uniform and hence the packing frustration gets minimized. For example, in a structure with four B-domains, if the four B-centers are arranged in a two-dimensional tetragonal or a square shape, the stretching of the majority blocks is much more nonuniform and induces larger packing frustration than that in the case of the four B-centers being in a regular tetrahedral shape.

Snapshots shown in Figure 1 include both stable and metastable structures, where each metastable structure is marked with a frame of dashed lines. It is noted that structures $S_2$, $S_3$, $S_8$ and $S_{11}$ are metastable in systems with $-\varepsilon_{AW}=\varepsilon_{BW}=1$ but they are stable in systems with $\varepsilon_{AW}=\varepsilon_{BW}=0$, while structures $S_5$ and $S_{13}$ are metastable in both of these two types of systems, indicating that the stability of some structures depends on the polymer-pore surface interactions. And the influence of the interactions on the self-assembled structures and stability are further investigated.

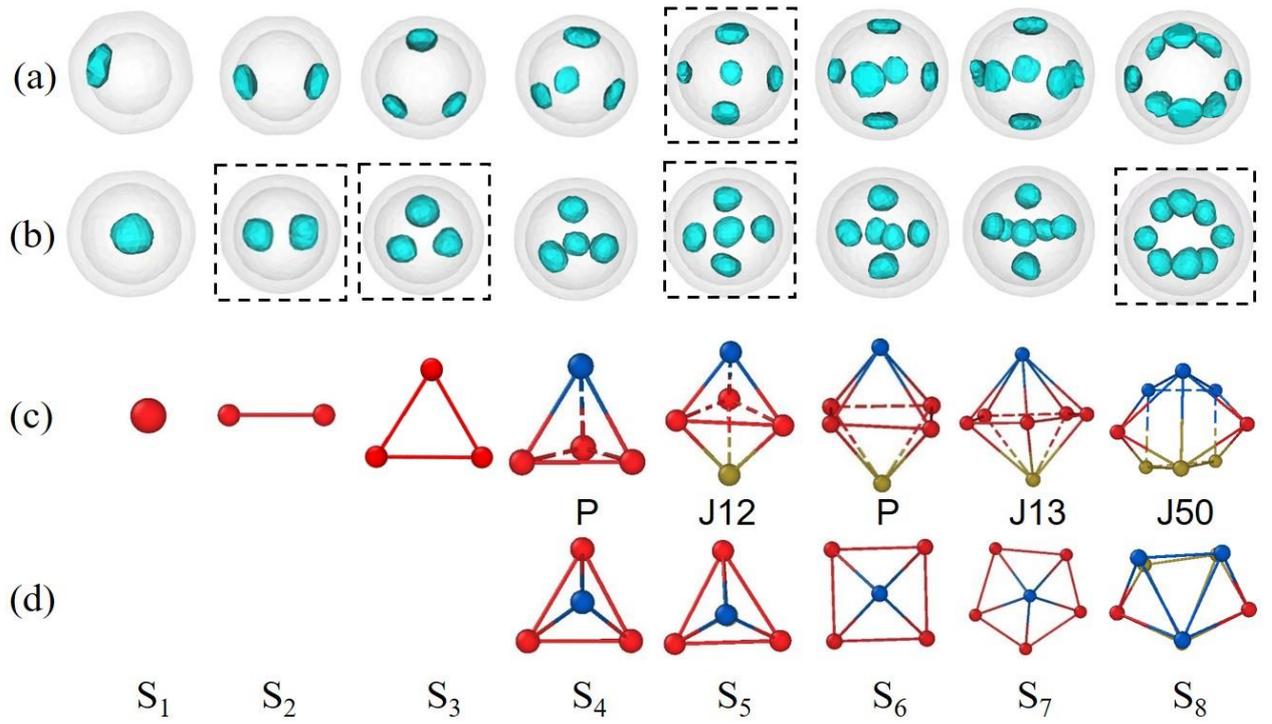

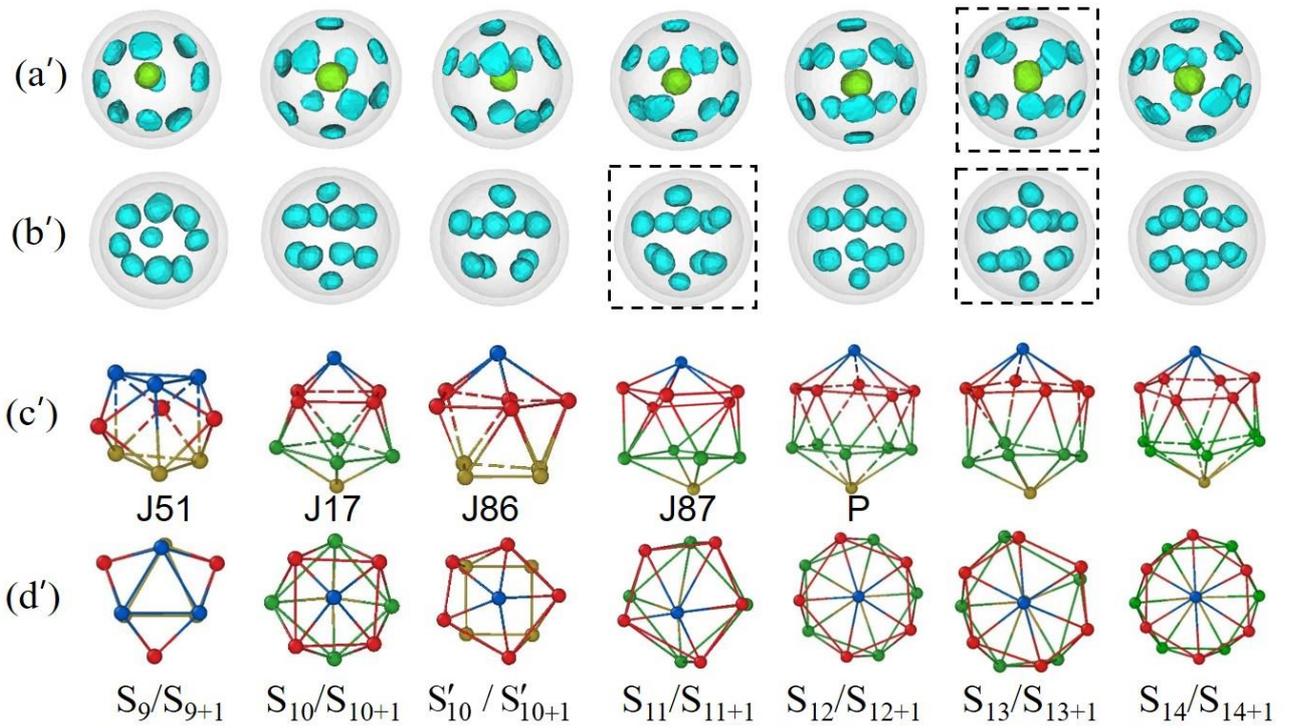

Figure 1. Typical snapshots of morphologies obtained at (a, a') $\varepsilon_{AW}=\varepsilon_{BW}=0$ and (b, b') $-\varepsilon_{AW}=\varepsilon_{BW}=1.0$. In (a, a', b, b'), isosurface contour plot of only B-domains is shown. Structures $S_1$ to $S_{14}/S_{14+1}$ are obtained at pore radius $R = 4, 6, 7, 8, 9, 10, 11, 11, 12, 13, 13, 14, 15, 15$, respectively in (a, a'), and at $R = 8, 9, 9, 10, 10, 11, 12, 13, 13, 14, 14, 15, 16, 16$, respectively in (b, b'). In (a, a', b, b'), the outermost

B-domains are shown in blue while the B-domain at the pore center in green color. In (c, d) or (c', d'), each dot corresponds to the center of a B-domain shown in (b) or (b'), respectively, and lines are added to guide the eyes, where the dots and lines are assigned to different colors only for clarity. (c) and (d), and (c') and (d') are viewed in two perpendicular directions. Metastable structures are labeled by framing the snapshots with dashed lines in (a, b, a', b').

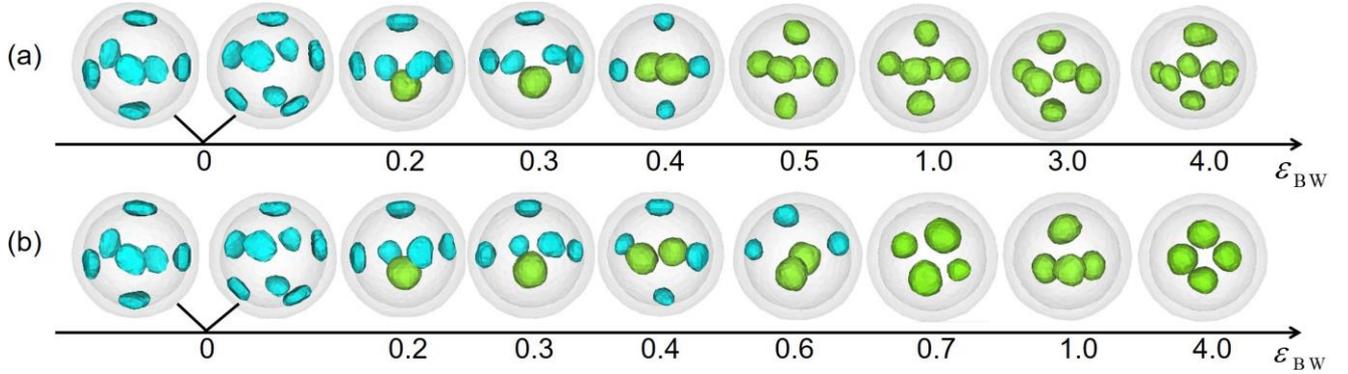

Figure 2. Typical snapshots of self-assembled morphologies as a function of $\varepsilon_{BW}$ for systems in the pore of radius of $R=11$ with the A-surface interaction being $\varepsilon_{AW}=-\varepsilon_{BW}$ in (a) and $\varepsilon_{AW}=0$ in (b). The B-domains attached to the pore surface are shown in blue and those away from the surface are shown in green.

Typical snapshots of self-assembled morphologies as a function of $\varepsilon_{BW}$ at a fixed pore radius of $R=11$ are shown in Figure 2 to illustrate the effect of the pore surface interactions on the self-assembled morphologies. Here results from two types of cases are presented with the only difference being in $\varepsilon_{AW}$, i.e., $\varepsilon_{AW}=-\varepsilon_{BW}$ in type $\alpha$ while $\varepsilon_{AW}=0$ in type $\beta$. It is noted that the B-domains observed in Figure 2 can be classified into two categories according to their differences in domain shape and position. For the sake of clarity, we denote the B-domains being in nearly lens-shape and attached to the pore surface (in blue color in Figure 2) as category 1, and those being in spherical shape and far away from the pore surface (in green color) as category 2. As shown in Figure 2a for the type $\alpha$ case, the number and the category of B-domains change with increasing $\varepsilon_{BW}$ in the following way: seven or eight B-domains all belonging to category 1 at $\varepsilon_{BW}=0$ → six B-domains with five of them belonging to category 1 while the other one to category 2 at $\varepsilon_{BW}=0.2$–0.3 → six B-domains with four of them belonging to category 1 while the other two to category 2($\varepsilon_{BW}=0.4$) → six or seven B-spheres all belonging to category 2 ($\varepsilon_{BW}=0.5$–3.0 or $\varepsilon_{BW}=4.0$). That is, the number of the B-domains shows a nonmonotonic change with $\varepsilon_{BW}$. In the type $\beta$ case, as shown in Figure 2b, the variation of domain shape with $\varepsilon_{BW}$ is similar to that in the type $\alpha$ case when $\varepsilon_{BW}$=0–0.4. However, the number of B-domains continues to decrease to five and further to four when increasing $\varepsilon_{BW}$ to 0.6 and further to 0.7, respectively, and finally keeps in four when $\varepsilon_{BW} \geq 0.7$.

It is noted that B-domains belonging to two categories simultaneously occur in each system as indicated by green and blue colors, at $\varepsilon_{BW}$=0.2–0.4 in Figure 2a and $\varepsilon_{BW}$=0.2–0.6 in Figure 2b. For these structures with two categories of B-domains, we name them as $S_{i+j}$, with $i$ and $j$ being the number of the outermost and inner B-domains, respectively. Our further results show that such structures may also occur at other pore radii when $\varepsilon_{BW}$ is at the above-mentioned $\varepsilon_{BW}$ regions. Figure 3 shows typical snapshots of structures obtained in the type $\beta$ case at $\varepsilon_{AW}$=0 and $\varepsilon_{BW}$=0.5 as a function of pore radius $R$. As shown in Figure 3, with increasing $R$, the structure changes from $S_1$ ($R$=4–10) → $S_{3+2}/S_{4+2}$ ($R$=11) → $S_{5+2}$ ($R$=12) → $S_{4+4}/S_{5+3}$($R$=13) → $S_{5+4}/S_{4+5}$ ($R$=14) → $S_{5+5}/S_{4+6}$ ($R$=15) → $S_{6+6}/S_{5+7}$ ($R$=16) → $S_{6+8}$ ($R$=17). At very small pore radius of $R$=4–10, only one B-domain is formed in the system. It is interesting to notice that the B-domain is attached to the pore surface when $R$=4–5, while it is located at the pore center when $R$=6–10. With increasing $R$, the total number of B-domains increases and B-domains belonging to two categories (in blue and green colors) coexist in each system. With increasing $R$ from 11 to 17, the number of B-domains away from the surface (in green color) increases from 2 → 3 or 4 → 4 or 5 → 5 or 6 → 6 or 7 → 8. It should be mentioned that the structures $S_{i+j}$ shown in Figure 3 are different from the structures $S_k$ shown in Figure 1 even when $k=i+j$. For example, the structure $S_{4+2}$ is different from the structure $S_6$. Furthermore, structure $S_{i+j}$ usually constitutes a polyhedron without apparent symmetry due to the different distances from the B-centers belonging to two categories to the pore center. However, we do observe a polyhedron with apparent symmetry constituted from structure $S_{4+4}$, it is the triangular dodecahedron belonging to the Johnson solid J84, as shown in Figure 3 at $R$=13. Structure J84 has eight vertices, four of which are farther from the pore center than the other four, and the B-centers in structure $S_{4+4}$ just meet such a condition, whereas the eight B-centers in structure $S_8$, being of the same category and of the same distances to the pore center, cannot meet the condition, thus polyhedron J84 is not observed in structure $S_8$. Structure $S_{5+3}$, observed at $R$=13 as a degenerate structure with $S_{4+4}$, is a deformed J84, here called DJ84. Our simulation results show that structures $S_{i+j}$ obtained in the type $\alpha$ case at $-\varepsilon_{BW}=\varepsilon_{BW}$=0.4 are similar to those obtained in the type $\beta$ case shown in Figure 3.

The occurrence of two categories of B-domains in each system can be understood based on the following analysis. When the pore surface is neutral to the two blocks ($\varepsilon_{AW}=\varepsilon_{BW}=0$), the B-domains are attached to the pore surface due to the two reasons mentioned earlier. Thus there is a tendency (the first tendency) to make the B-domains attached to the pore surface. With increasing $\varepsilon_{BW}$, the repulsive interaction between the B-segments and the pore surface tends to make the B-domains away from the pore surface (the second tendency). Hence all the B-domains can be away from the pore surface only when $\varepsilon_{BW}$ is large enough so that the second tendency can offset or stronger than the first tendency. From Figure 2, it is noted that all the B-domains are away from the pore surface at $\varepsilon_{BW}=0.5$ and $0.7$ for the type $\alpha$ and $\beta$ case, respectively, while when $0<\varepsilon_{BW}<0.5$ (or $0<\varepsilon_{BW}<0.7$) for the type $\alpha$ (or $\beta$) case, competition between the two tendencies results in different structures as shown in Figure 3. At very small pore, the amount of block copolymers in a system is small, and hence only one B-domain is formed there. When $R=4$–$5$, $R$ is much smaller than the length of the solvophilic A-block, and hence the B-domain is attached to the pore surface so that the A-blocks can have relatively large space being less compressed. In this case, the punishment due to the repulsive interaction between the pore surface and B-blocks can be offset by the reduced entropy lose due to the less compressed A-blocks. When $R=6$, the amount of block copolymers in the system is about twice that when $R=4$–$5$. In this case, the system may form one big B-domain or two normal B-domains. If the system forms two normal B-domains the locations of them would be one of the following three patterns: 1. both of them are attached to the pore surface, i.e., the structure $S_2$ shown in Figure 1a; 2. both of them are away from the pore surface, i.e., the structure $S_2$ shown in Figure 1b; 3. one of them is attached to the pore surface, while the other is away from the pore surface. Pattern 1 would result in a high energy due to the increased contact area between the B-segment and the pore surface, while patterns 2 and 3 means that the distance between the two B-domains is less than the pore diameter, i.e., the space for A-blocks connected to the B-domains is less than that in a pore of $R=4$–$5$, which would result in strongly compressed A-blocks. On the other hand, the system may form one big B-domain. In this case, if the B-domain is attached to the pore surface, the A-blocks emitting from the surface of the B-domain would be too crowded since that the A-blocks are only distributed at the inner side of the B-domain surface (the outer side is attached to the pore surface). The crowded distribution of A-segments reduces the contact between A-segments and solvent, and hence it is energetically unfavorable. On the other hand, if the big B-domain is located at the pore center, it can remove the contact between the B-segment and the pore surface, and can make the space for each A-block relatively large, and can make the A-blocks not too crowded; and therefore, it is the best choice in this case for system with $R=6$. For systems with $R=7$–$10$, the amount of block copolymers in the system is more than three times that of when $R=4$–$5$. In this case, the system may form one big B-domain or several normal B-domains. Based on an analysis similar to that for the system with $R=6$ it can also be deduced that forming one big B-domain located at the pore center is the best choice for these systems. As the pore is further larger to $R=11$, if the system remains forming one big B-domain located at the pore center, the stretching of both B-blocks and A-blocks will be strong, which is entropically unfavorable. And simultaneously A-blocks would be much crowded near the B-domain surface, which will reduce the contact between A-segments and solvent. Hence forming one big B-domain located is energetically unfavorable at $R=11$ as analyzed above. On the other hand, if the system forms several normal B-domains with some of them being attached to the pore surface while others being away from the pore surface, which may allocate the space effectively so that the stretching of both B-blocks and A-blocks and the crowding of A-blocks are all relieved, and the resulting favorable in both entropic and energetical side can offset the punishment due to the increased unfavorable contact between the B-segments and the pore surface. This is the reason that two categories of B-domains are formed for the system with $R \geq 11$ as shown in Figure 3. On the other hand, at a given $\varepsilon_{BW}$, the interaction $\varepsilon_{AW}=-\varepsilon_{BW}$ used in the type $\alpha$ case is equivalent to a stronger surface selectivity than that used in the type $\beta$ case with $\varepsilon_{AW}=0$, which is the reason that a relatively smaller $\varepsilon_{BW}$ value is needed in the type $\alpha$ case to make all the B-domains away from the pore surface than that needed in the type $\beta$ case.

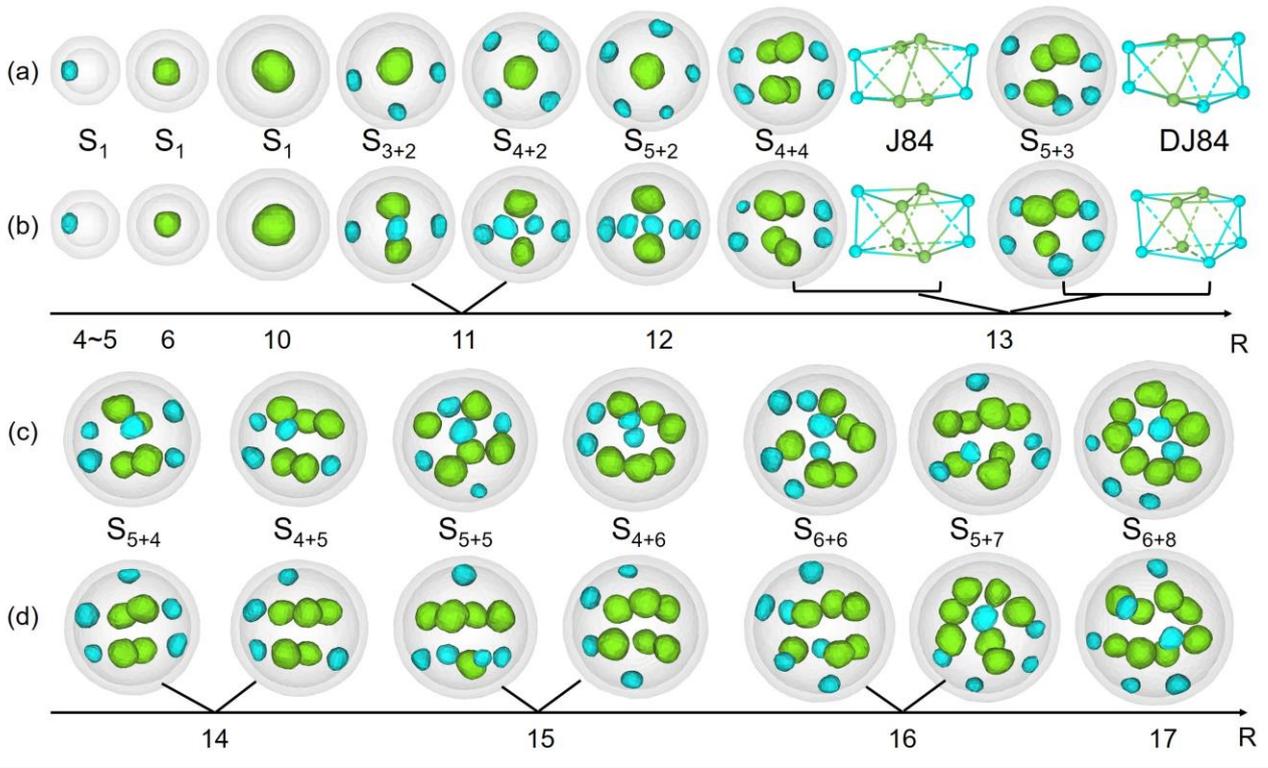

Figure 3. Typical snapshots of self-assembled morphologies as a function of $R$ for system with block-surface interactions being $\varepsilon_{AW}=0$ and $\varepsilon_{BW}=0.5$. (a) and (b), and (c) and (d) are viewed from two different directions. The B-domains attached to the pore surface are shown in blue while those away from the surface are shown in green. The B-centers connected by lines are shown for structures $S_{4+4}$ and $S_{5+3}$ formed at $R=13$ to illustrate the structures J84 and "DJ84".

## 3.2 Phase diagrams

We further construct two phase diagrams for the type $\alpha$ and $\beta$ case as shown in Figure 4a and 4b, respectively, for the stable region of the obtained structures $S_i$ and $S_{i+j}$ as a function of the pore radius $R$ and the interaction strength of the pore surface with B-blocks $\varepsilon_{BW}$. Note that the two types of cases are the same at $\varepsilon_{AW}=\varepsilon_{BW}=0$ (neutral surface), and similar morphologies including two categories of B-domains are obtained when $\varepsilon_{BW}$ is at $\varepsilon_{BW}=0.2$–0.4 and at $\varepsilon_{BW}=0.2$–0.6 as shown in Figure 2a and 2b, respectively. Thus we now focus on the different phase behavior between the two types of cases when each system forms only one category of B-domains which are away from the pore surface, i.e., when $\varepsilon_{BW}\geq 0.5$ in the type $\alpha$ case and $\varepsilon_{BW}\geq 1.0$ in the type $\beta$ case. As the polyhedrons constituted from structures $S_i$ with $i \geq 15$ do not have apparent symmetry, we mainly focus on the stable region of structures $S_1$–$S_{14}$, where three main differences between the two types of cases are found as listed in the followings.

1. As shown in Figure 4a for the type $\alpha$ case, with increasing $R$ and $\varepsilon_{BW}$, each of the structures $S_1$–$S_{14}$ has a stable region in the phase diagram, though the stable region of some structures is very small. It should be noted that the metastable structures marked with a frame of dashed lines shown in Figure 1, are stable at other interaction strengths as shown in the phase diagram in Figure 4a. The structures with the stable regions ranked from largest to smallest are approximately in the following order: $S_1$, $S_4$, $S_6$, $S_9$, $S_{12}$ → $S_2$, $S_3$, $S_7$, $S_8$, $S_{10}$ → $S_5$, $S_{11}$, $S_{14}$ → $S_{13}$. It is noted that structures $S_1$, $S_4$, $S_6$, $S_9$, $S_{12}$ occur in the whole $\varepsilon_{BW}$ range, while other structures only occur at some $\varepsilon_{BW}$ values. As shown in Figure 4b for the type $\beta$ case, structures $S_1$, $S_4$, $S_6$, $S_7$, $S_8$, $S_9$, $S_{12}$ and $S_{14}$ keep stable when $\varepsilon_{BW} \geq 1.0$ and $S_{10}$ is stable only when $1.0 \leq \varepsilon_{BW} \leq 1.5$, while structures $S_2$, $S_3$, $S_5$, $S_{11}$ and $S_{13}$ are not stable when $\varepsilon_{BW} \geq 1.0$. It is obviously that the stable structures in the type $\alpha$ case are much richer than those in the type $\beta$ case. On the other hand, it is noted that structures $S_1$, $S_4$, $S_6$, $S_9$, and $S_{12}$ have relatively larger stable regions in the diagram, which is the same as that in the type $\alpha$ case. It is interesting to notice that these structures of relatively larger stable regions correspond to the three observed Platonic solids of regular tetrahedron ($S_4$), octahedron ($S_6$) and icosahedron ($S_{12}$) and one Johnson solid of J51 ($S_9$).

2. In the type $\alpha$ case, a nonmonotonic change of the number of the B-domains in the formed structures is found when $\varepsilon_{BW}$ is increased from 0 to 4.0 at a fixed pore radius $R$, e.g., the sequences of $S_3 \to S_1 \to S_2 \to S_3 \to S_4$, $S_6 \to S_4 \to S_5 \to S_6$, or $S_{12} \to S_9 \to S_{10} \to S_{12}$ are found when fixing the value of $R$ at 7, 10, or 14, respectively. That is, at a given $R$, the number of the B-domains has a minimum value at $\varepsilon_{BW} \approx 0.5$, and after that minimum value, it increases with $\varepsilon_{BW}$ in the listed $\varepsilon_{BW}$ range. While in the type $\beta$ case, at a fixed $R$, the number of the B-domains also decreases from $\varepsilon_{BW}=0$ to 0.7, whereas it keeps almost unchanged with further increasing $\varepsilon_{BW}$ when $\varepsilon_{BW} \geq 1.0$. This could be understood by the difference of the value of $\varepsilon_{AW}$ used in the types $\alpha$ and $\beta$ cases. As mentioned earlier, all the B-domains are away from the pore surface at $\varepsilon_{BW}=0.5$ and 0.7 for the type $\alpha$ and $\beta$ case, respectively. On the other hand, with further increasing $\varepsilon_{BW}$, the attractions between the pore surface and A-blocks are strengthened in the type $\alpha$ case and hence the A blocks tend to move towards the pore surface, which leads to that the B-blocks, connected with the A-blocks covalently, are pulled towards the pore surface. Therefore, the effective pore radius for the outermost B-domains should increase with further increasing $\varepsilon_{BW}$. On the other hand, increasing the number of B-domains is helpful for more contact between A-segment and pore surface, which is energetically favorable in the type $\alpha$ case. Hence the number of B-domains increases with $\varepsilon_{BW}$ when $\varepsilon_{BW}>0.5$, which is the reason of the nonmonotonic change of the number of the B-domains with $\varepsilon_{BW}$ in the formed structures at a fixing pore radius $R$ in the type $\alpha$ case. While in the type $\beta$ case, the value of $\varepsilon_{AW}$ is always fixed at zero. When $\varepsilon_{BW}$ is large enough, i.e., $\varepsilon_{BW}\geq1.0$, further increasing $\varepsilon_{BW}$ would not affect the behavior of B-blocks since that the pore surface and B-domains are separated by A-blocks and solvent. Therefore, the number of B-domains keeps unchanged with increasing $\varepsilon_{BW}$ when $\varepsilon_{BW}\geq1.0$ at a fixing $R$ in the type $\beta$ case.

3. In the type $\beta$ case, to form a structure with the same number of B-domains as that in the type $\alpha$ case, a larger pore radius $R$ is needed. For example, the structure $S_4$ appears at $R=7–10$ depending on $\varepsilon_{BW}$ in the type $\alpha$ case, while it appears at a larger pore radius of $R=11$ in the type $\beta$ case. This further indicates that the effective pore radius for the outermost B-domains should be different in the studied two types of cases.

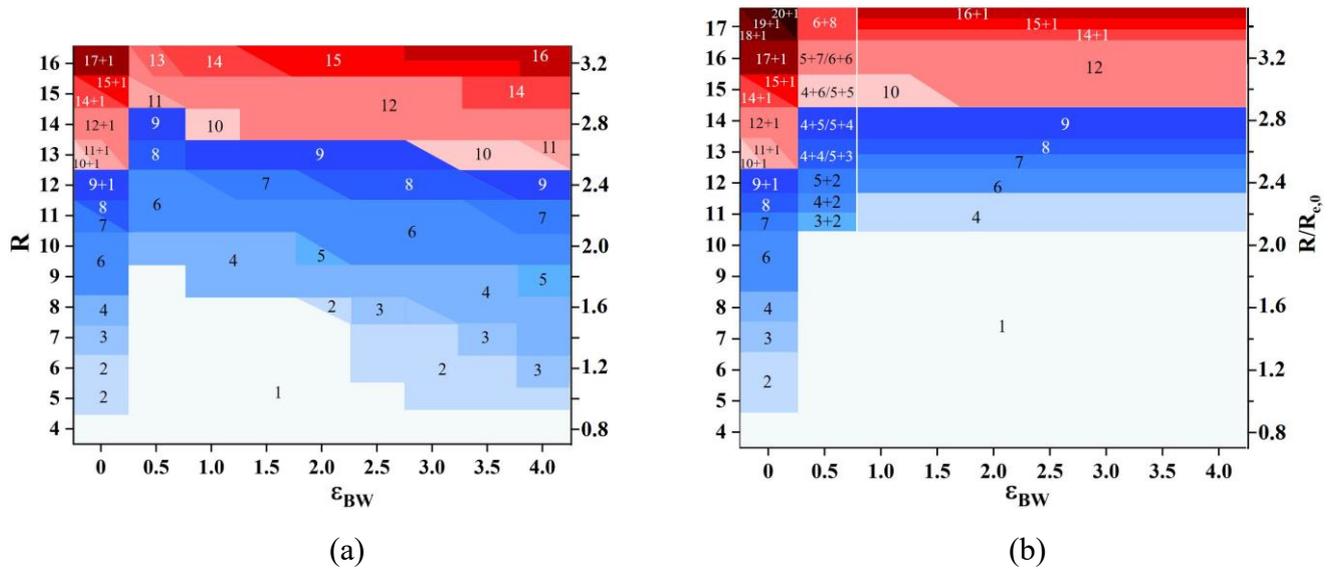

(a)  (b)

Figure 4. Phase diagrams showing the number of solvophobic B-domains in the obtained stable structures as a function of the pore radius $R$ (or $R/R_{e,0}$, where $R_{e,0}$ being the root of the mean-square end-to-end distance of the corresponding ideal chain) and the interaction parameter $\varepsilon_{BW}$ in a step of 0.5. The corresponding typical structures are shown in Figures 1–3. The A-surface interactions are $\varepsilon_{AW}=-\varepsilon_{BW}$ in (a) and $\varepsilon_{AW}=0$ in (b). For clarify, the background color varies from light blue to dark blue to light red and further to dark red showing that the number of B-domains varies from small to large.

## 3.3 Quantitative comparison of the details of structures between the two types of cases

We calculate some quantities to provide quantitative information for structures obtained in the two types of cases. The above analysis suggests that the effective pore radius for the B-domains should be different in the two types of cases for a given $R$. We estimate the effective pore radius $R_{\text{eff}}$ through calculating the density profiles of I-segments $P_I(r)$ (I=A, B) along the radical direction of the confining pore. Here $P_I(r)=n_I(r)/M(r)$ with $n_I(r)$ and $M(r)$ being the number of I-segments and the number of lattice sites in the spherical shell of thickness 1 at a distance $r$ from the pore center. Figure 5 shows the density profiles of A- and B-segments for systems of $R=11$, corresponding to the structures shown in Figure 2. As shown in Figures 5a and 5c for the two types of cases, a $P_B(r)$ curve has one peak or two peaks corresponding to systems with one category or two categories of the B-domains, respectively. Without loss of generality, the effective pore radius $R_{\text{eff}}$ for the B-domains is defined as the peak position of the $P_B(r)$ curve. It is obviously that when there is only one category of B-domains and they are away from the pore surface, the $R_{\text{eff}}$ in the type $\alpha$ case is larger than that in the type $\beta$ case, i.e., $R_{\text{eff}} \approx 7$ in the type $\alpha$ case at $\varepsilon_{BW} \geq 0.5$ and $R_{\text{eff}} \approx 5$ in the type $\beta$ case at $\varepsilon_{BW} \geq 0.7$ as indicated by the peak position of the $P_B(r)$ curves shown in Figure 5a and 5c. This difference is due to the different $\varepsilon_{AW}$ values used in the two types of cases. Most of the A-segments are attached to the pore surface due to the attractive interactions between the A-blocks and surface in the type $\alpha$ case, as indicated by the $P_A(r)$ curves where a peak occurs at $r=R=11$ when $\varepsilon_{BW} \geq 0.5$ shown in Figure 5b. Considering the size of each B-domain and the stretching of the solvophilic A-blocks, $R_{\text{eff}} \approx 7$ means that the location of the B-domains is quite near the pore surface for a pore of radius 11. While in the type $\beta$ case as shown in Figure 5d, a $P_A(r)$ curve has a small peak at $r=R-1$ and a much lower value at $r=R$, indicating that much more solvent molecules than A-segments are located at the pore surface when $\varepsilon_{BW} \geq 0.7$. This is because in the type $\beta$ case the pore surface has no energetic preference to A-segment over solvent as $\varepsilon_{AW}=\varepsilon_{SW}=0$, while more solvent molecules located at the pore surface can reduce the entropy loss of the polymer. In this case, the peak of a $P_B(r)$ curve, is located at $r=5$, i.e., at about the center between the pore center and the outermost layer mostly occupied by solvent, which makes the space being about the same for the A-blocks emitting from a B-domain surface stretching to both the pore center and the pore surface directions.

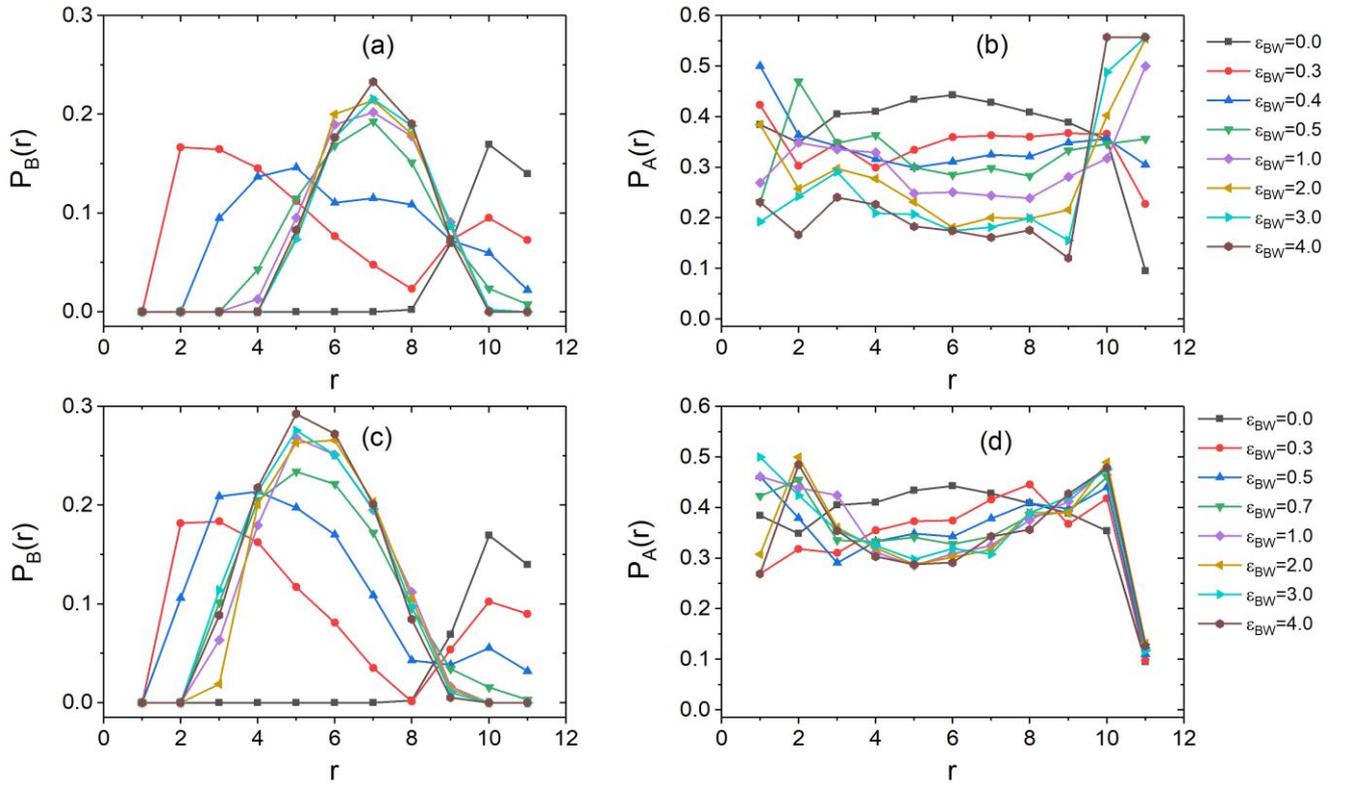

Figure 5. Density profiles of A-segments $P_A(r)$ and B-segments $P_B(r)$ along the radical direction of the confining pore of radius $R=11$. (a, b) Type $\alpha$ case and (c, d) type $\beta$ case.

We also calculate other three types of quantities to provide information of morphologies or chains, including nonsphericity of each B-domain ($\kappa$), the index of polydispersity of the size of each B-domain (*pdi*) and the normalized mean-square end-to-end distance of AB diblock chains ($R_e^2$), A-blocks ($R_{e,A}^2$) and B-blocks ($R_{e,B}^2$). These quantities as a function of $\varepsilon_{BW}$ are shown in Figure 6, where empty and filled symbols correspond to the type $\alpha$ and $\beta$ case, respectively. From Figure 6a, it is noted that in both types of $\alpha$ and $\beta$ cases, the $\kappa$ curve has a maximum value at $\varepsilon_{BW}=0$, indicating that the shape of each B-domain is deviating from a sphere, which is because that the B-domains are attached to the pore surface and in lens-shape at $\varepsilon_{BW}=0$. With increasing $\varepsilon_{BW}$, each $\kappa$ curve decreases rapidly till $\varepsilon_{BW}=0.5$ or 0.7 for the type $\alpha$ and $\beta$ case, respectively, where $\kappa$ reaches a minimum value of ~0.02. This is because at these small $\varepsilon_{BW}$ region, each system includes two categories of B-domains, lens-shaped and nearly spherical, as indicated by green and blue colors in Figure 2, and the proportion of the latter increases with $\varepsilon_{BW}$ and becomes 100% at $\varepsilon_{BW}=0.5$ or 0.7 for the type $\alpha$ and $\beta$ case, respectively. With further increasing $\varepsilon_{BW}$, $\kappa$ keeps at the minimum value almost unchanged in the type $\beta$ case, indicating that the B-domains are in nearly spherical shape. While in the type $\alpha$ case, however, $\kappa$ increases slightly with $\varepsilon_{BW}$ in the range of $\varepsilon_{BW}$=1.5–4.0, indicating that the strong A-surface attractions induce a slight deformation of the shape of each B-domain.

As shown in Figure 6b, the index of polydispersity of the size of each B-domain (*pdi*) has a much larger value of ~1.3 at $\varepsilon_{BW}$=0.2–0.4 and of 1.3–1.5 at $\varepsilon_{BW}$=0.2–0.6 for the type $\alpha$ and $\beta$ case, respectively. These $\varepsilon_{BW}$ regions just correspond to the structures with two categories of B-domains in the system. In the other $\varepsilon_{BW}$ regions, *pdi* keeps at a value of close to 1.0 and almost unchanged in the type $\beta$ case. While in the type $\alpha$ case, it also keeps an almost unchanged value of ~1.0 at $\varepsilon_{BW}$=1.0–3.0, however, it increases slightly when $\varepsilon_{BW}$≥3.5, indicating that the strong A-surface attractions in the type $\alpha$ case can induce a slight difference in the sizes of B-domains. Figure 6c shows the variation of $R_e^2$, $R_{e,A}^2$ and $R_{e,B}^2$ for chains in the self-assembled structures as a function of $\varepsilon_{BW}$. As the length of the B-block is much shorter than that of the A-block, the stretching of the AB chains (characterized by $R_e^2$) is mainly dominated by the stretching of the A-blocks (characterized by $R_{e,A}^2$). This can be clearly seen from Figure 6c where the variations of $R_e^2$ and $R_{e,A}^2$ are almost synchronously. The value of $R_{e,A}^2 \approx 1.7$ as shown in Figure 6c at $\varepsilon_{BW}$=0, indicates that the A-blocks are strongly stretched there. This is because when $\varepsilon_{AW}=\varepsilon_{BW}=0$, the shorter B-blocks form lens-shaped domains attached to the neutral pore surface, and hence the A-blocks are only distributed at the inner side of each B-domain, which results in the large $R_{e,A}^2$ value. With increasing $\varepsilon_{BW}$ from 0 to ~0.5, the stretching of the A-blocks is slightly relieved as shown in Figure 6c. This is because that with increasing $\varepsilon_{BW}$, more and more B-domains move away from the pore surface, and hence the A-blocks are distributed at all sides of each B-domain, which relieves the stretching of the A-blocks. With further increasing $\varepsilon_{BW}$, $R_{e,A}^2$ maintains almost unchanged in the type $\beta$ case, while in the type $\alpha$ case $R_{e,A}^2$ further decreases slightly with increasing $\varepsilon_{BW}$ when $\varepsilon_{BW}$≥3.5 since that with increasing $\varepsilon_{BW}$, the attraction between A-segments and the pore surface increases and hence more A-segments are concentrated near the pore surface (as indicated in Figure 5b), which reduces the contact between A-segment and solvent, and hence reduces the stretching of the A-blocks. On the other hand, the increased number of B-domains with increasing $\varepsilon_{BW}$ when $\varepsilon_{BW}$≥3.5 (as shown in Figure 4a) can also reduce the stretching of A-blocks.

In short summary, each B-domain is in a more close to spherical shape at $\varepsilon_{BW}$=0.5–1.0 in the type $\alpha$ case and $\varepsilon_{BW}$≥0.7 in the type $\beta$ case; the sizes of B-domains are uniform except that when there are two categories of B-domains (attached to and away from the pore surface) in the system or the attraction between A-segments and the pore surface is strong; and the stretching of the A-blocks is strongest when the B-blocks are attached to the pore surface and it is somewhat relieved when the B-blocks are away from the pore surface or the number of B-domains is increased.

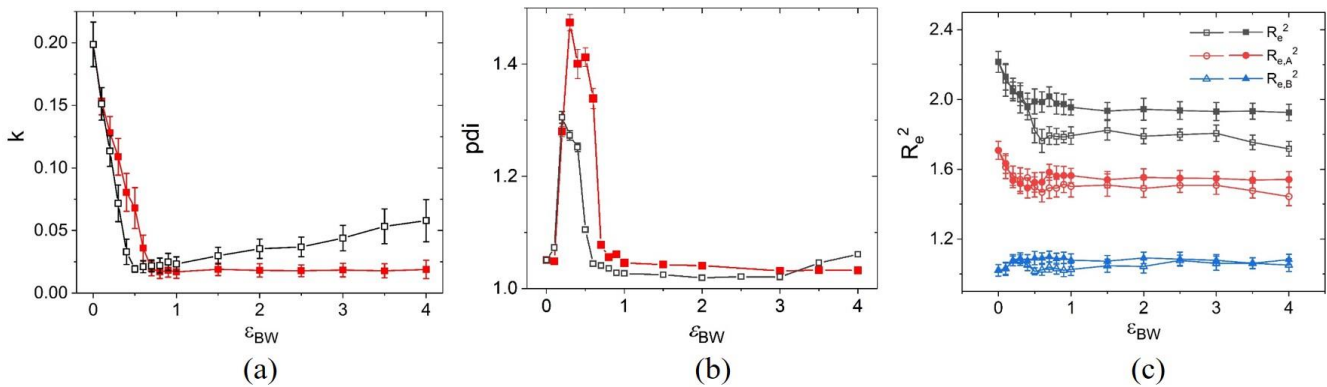

Figure 6. (a) Nonsphericity of each B-domain, (b) index of polydispersity of the size of each B-domain and (c) normalized mean-square end-to-end distance of AB diblock chains, A-blocks and B-blocks in the self-assembled structures as a function of $\varepsilon_{BW}$ for systems of $R$=11, where empty and filled symbols correspond to type $\alpha$ and $\beta$ case, respectively.

### 3.4 Effect of segment concentration on the self-assembled morphologies

Figure 7a shows the typical snapshots of self-assembled morphologies obtained by varying the copolymer segment concentration $C_p$ at a fixed pore radius of $R$=12 and fixed block-surface interactions of $-\varepsilon_{AW}=\varepsilon_{BW}$=1.0, and quantities calculated from these morphologies varying with $C_p$ are presented in Figure 7(b, c). As shown in Figure 7a, the B-domains always are in spherical up to $C_p$=0.7 (a value lower than that in unconfined system where up to $C_p$=0.85), and the number of the B-domains keeps six independent of $C_p$. Figure 7a also shows that the size of each B-domain increases with increasing $C_p$. The curve in Figure 7b shows the variation of relative distance between the six B-centers, characterized by the mean-square radius of gyration of the six B-centers $R_{g,c}^2$. The rapid increase of $R_{g,c}^2$ with decreasing $C_p$ indicates that the six B-centers are further away from each other with decreasing $C_p$. This can be understood based on the following analysis. With decreasing $C_p$, the amount of A-blocks in a system decreases, so the average size of each B-domain decreases as the number of the B-domains is not changed. As the outermost layer (at $r=R$) is occupied by the A-segment and the B-domain is close to the A-segments, so the distance of each B-center to the pore surface is about the radius of the B-domain plus 1. Hence with the decrease of the B-domain size, the B-centers are gradually close to the pore surface. That is, with decreasing $C_p$, the B-centers are on a spherical surface of gradually increased radius, therefore, the mean-square radius of gyration of the six B-centers $R_{g,c}^2$ increases with decreasing $C_p$. Figure 7c shows that the stretching of AB diblock chains, mainly dominated by the stretching of the longer A-blocks as mentioned before, increases rapidly with decreasing $C_p$. It can be deduced that with decreasing $C_p$, the A-blocks are more and more swollen by the increased amount of solvent molecules, thus they tend to be more and more stretched with decreasing $C_p$.

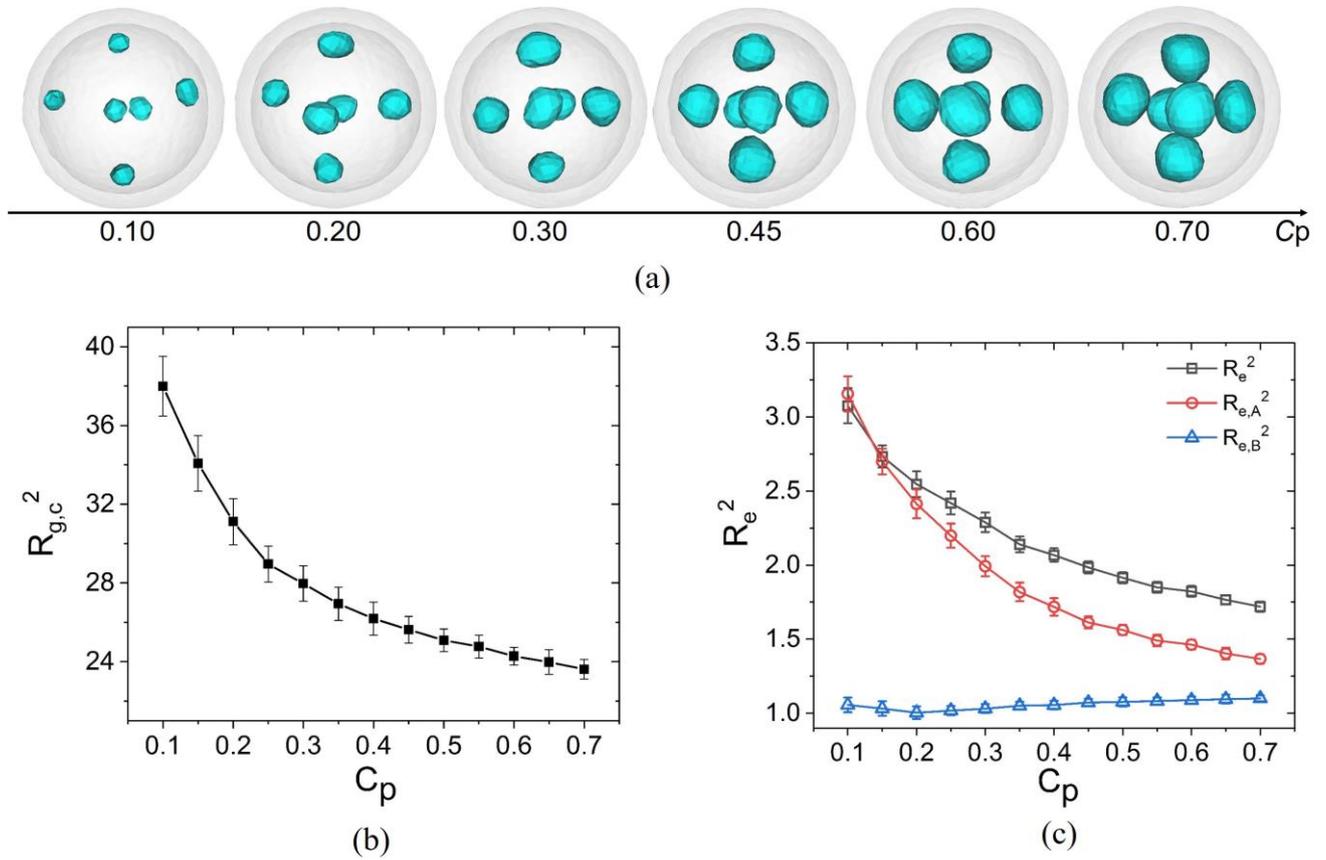

Figure 7. Segment concentration ($C_p$) dependence of (a) typical snapshots of self-assembled morphologies, (b) mean-square radius of gyration of the centers of B-domains, and (c) normalized mean-square end-to-end distance of AB diblock chains, A-blocks and B-blocks in the self-assembled structures in systems of $R=12$ and the block-surface interaction of $-\varepsilon_{AW}=\varepsilon_{BW}=1.0$. The spherical B-domains in (a) are shown in blue color.

Our further simulations show that the number of the B-domains may change little with $C_p$ at other fixed pore radii. On the other hand, varying the volume fraction of one block or the selectivity of the solvent, can affect the $C_p$ region for forming spherical structures, just like that observed by Lodge et al. earlier[32]. Nevertheless, the phase behavior obtained in the present study should be general as long as the solution-system forms spherical structures, regardless of the values of $f_B$, solvent selectivity and $C_p$.

## 3.5 Comparisons with related works

To our knowledge, there are no reports on studies of the self-assembly of sphere-forming amphiphilic diblock copolymer solutions confined in spherical nanopores both on the experimental and theoretical/simulation front. However, our simulation results can be compared with that of related study of self-assembly of sphere-forming diblock copolymer melts confined in spherical nanopores by Zhao et al.[31] using self-consistent field theory. The main differences between our system/results and theirs are summarized in the followings. (1) Our model system is in solution-state while theirs is in melt. (2) We focus on the neutral pore surface and surface selective for the longer blocks while they focus on pore surface selective for the shorter blocks. (3) Our phase diagrams are constructed through varying the pore radius and the surface-block interaction at a fixed volume fraction of B-block being 3/16, while their phase diagram is constructed as a function of the pore radius and the volume fraction of A-blocks at fixed surface-block interactions. (4) From our phase diagram each of structures $S_1$–$S_{14}$, being of apparent symmetry, has a stable region in the type $\alpha$ case, while in the type $\beta$ case, structures $S_2$, $S_3$, $S_5$, $S_{11}$ and $S_{13}$ are not observed as stable when $\varepsilon_{BW}>0$. From their phase diagram, structures $S_5$ and $S_{11}$ do not have a stable region among the twelve structures $S_1$–$S_{12}$ they studied. (5) Structure $S_8$ is the biaugmented triangular prism belonging to J50 in the Johnson solids in our results, while in their results it is the triangular dodecahedron belonging to J84 in the Johnson solids. The shape of the J50 obtained in our simulation is closer to a sphere comparing to that of J84 (which has a larger aspect ratio). They did not consider J50 as a candidate structure in their study. On the other hand, J84 is obtained in our study from structure $S_{4+4}$ which is a structure with two categories of B-domains. (6) We obtain two degenerate structures for structure $S_{10}$, they are J17 and J86, while in their study only J17 was considered as a candidate structure for $S_{10}$ and it was found to be stable. The differences in results between theirs and ours should be mainly due to the differences in model/parameters summarized above in differences (1–3). Another important factor that may result in differences between their results and ours is the fluctuation of the system which is considered automatically in our simulations while was not considered in their self-consistent field theory calculation. Fluctuation is more important in solution-state system than that in melt system. On the other hand, some similar results are found between theirs and ours, for example, the five structures $S_1$, $S_4$, $S_6$, $S_9$, and $S_{12}$ have much larger stable regions in the phase diagram than others. It should be due to the higher symmetry of these five structures that makes all chains stretching uniformly. In addition, their study showed that the stability order of the three candidate structures (triangular dodecahedron, anticube, cube) for $S_8$ is triangular dodecahedron > anticube > cube, which supports our conclusion that the system prefers structures with more triangular faces, where the shape of the structure is closer to a sphere, and hence the stretching of the majority blocks is relatively uniform.

## 4 Conclusion

Self-assembly of amphiphilic sphere-forming AB diblock copolymer solutions confined in a spherical nanopore is studied using simulated annealing technique. Two types of cases of different pore-surface/copolymer interactions are studied: ($\alpha$) the surface being neutral or attractive to the solvophilic A-blocks and simultaneously repulsive to the solvophobic B-blocks, i.e., $-\varepsilon_{AW}=\varepsilon_{BW}\geq0$, ($\beta$) the surface being neutral or repulsive to the solvophobic B-blocks, i.e., $\varepsilon_{BW}\geq0$ while keeps $\varepsilon_{AW}=0$ unchanged. Structures with various number of solvophobic domains are obtained and named as $S_i$ or $S_{i+j}$, with $i$ and $j$ being the number of solvophobic domains located at the outermost and the inner of the confining pore, respectively. Phase diagrams are constructed as a function of the pore radius $R$ and the interaction parameter $\varepsilon_{BW}$ for the two types of cases. It is noted that depending on $R$, structures $S_1$, $S_4$, $S_6$, $S_9$, and $S_{12}$ occur in the whole $\varepsilon_{BW}$ range, while other structures only occur at some $\varepsilon_{BW}$ values or even do not occur in the type $\beta$ case when $\varepsilon_{BW}>0$. Furthermore, polyhedrons constituted by connecting the centers of the $i$ solvophobic domains in structure $S_i$ or $S_{i+1}$, are usually with high symmetry when $4\leq i \leq 14$, and especially three Platonic solids and seven Johnson solids are identified among them. These polyhedrons are usually with all or most of their faces being in triangular shape which makes the shape of the polyhedron being closer to a sphere. Regular tetrahedron, octahedron and icosahedron, constituted from structures $S_4$, $S_6$ and $S_{12}$, respectively, being composed of triangular faces and of relatively larger stable regions in the phase diagrams, are identified out of the five Platonic solids, while cube and regular dodecahedron, being not composed of triangular faces, are not observed here. At a fixed pore radius $R$, a nonmonotonic change of the number of B-domains with $\varepsilon_{BW}$ is observed in systems in the type $\alpha$ case. While in the type $\beta$ case, this number keeps almost unchanged in systems with one-category of B-domains when $\varepsilon_{BW} \geq 0.7$. When $\varepsilon_{BW}$ is relatively small, structures $S_{i+j}$ (with two-categories of B-domains) are observed and polyhedrons constituted by connecting the centers of the $i+j$ solvophobic domains in structure $S_{i+j}$ usually have no apparent symmetry in both types of $\alpha$ and $\beta$ cases. An exception is that, the $J_{84}$, the eighth Johnson solid in our study, is identified by connecting the centers of the eight solvophobic domains in $S_{4+4}$.

Our study shows that at a given $R$, the effective pore radius of the B-domains in the type $\beta$ case is smaller than that in the type $\alpha$ case when $\varepsilon_{BW} \geq 0.7$, and the size of each B-domain is uniform in systems with one category of B-domains. Furthermore, the solvophilic A-blocks have a larger stretching when all of the B-domains are attached to the pore surface in the case of neutral pore or when the segment concentration of the system is low.

The phase behavior obtained in the present study should be general as long as the solution-system forms spherical structures, regardless of the copolymer composition and concentration, as well as solvent selectivity. Our simulation results may provide guidelines for obtaining polyhedral-shaped structures from self-assembly of amphiphilic diblock copolymer solutions.

## Acknowledgements

This work was supported by the National Natural Science Foundation of China (21829301, 21774066 and 22173051), Fundamental Research Funds for the Central Universities, Nankai University (63221053), and Startup Funds for scholars of Nankai University, which is gratefully acknowledged.

References

(1) Bates, F. S.; Fredrickson, G. H. Block Copolymers—Designer Soft Materials. *Phys. Today* **1999**, *52* (2), 32. https://doi.org/10.1063/1.882522.

(2) Hamley, I. W. *The Physics of Block Copolymers*; Oxford University Press: Oxford, U.K., 1998.

(3) Russell, T. P.; Chai, Y. 50th Anniversary Perspective: Putting the Squeeze on Polymers: A Perspective on Polymer Thin Films and Interfaces. *Macromolecules* **2017**, *50* (12), 4597–4609. https://doi.org/10.1021/acs.macromol.7b00418.

(4) Schacher, F. H.; Rupar, P. A.; Manners, I. Functional Block Copolymers: Nanostructured

Materials with Emerging Applications. *Angew. Chem., Int. Ed.* **2012**, *51* (32), 2–25. https://doi.org/10.1002/anie.201200310.

(5)  Shi, A. C.; Li, B. Self-Assembly of Diblock Copolymers under Confinement. *Soft Matter* **2013**, *9* (5), 1398–1413. https://doi.org/10.1039/c2sm27031e.

(6)  Shin, J. J.; Kim, E. J.; Ku, K. H.; Lee, Y. J.; Hawker, C. J.; Kim, B. J. 100th Anniversary of Macromolecular Science Viewpoint: Block Copolymer Particles: Tuning Shape, Interfaces, and Morphology. *ACS Macro Lett.* **2020**, *9* (3), 306–317. https://doi.org/10.1021/acsmacrolett.0c00020.

(7)  Fasolka, M. J.; Mayes, A. M. Block Copolymer Thin Films: Physics and Applications. *Annu. Rev. Mater. Res.* **2001**, *31* (1), 323–355. https://doi.org/10.1146/annurev.matsci.31.1.323.

(8)  Segalman, R. A. Patterning with Block Copolymer Thin Films. *Mater. Sci. Eng. R Reports* **2005**, *48* (6), 191–226. https://doi.org/10.1016/j.mser.2004.12.003.

(9)  Cheng, J. Y.; Ross, C. A.; Smith, H. I.; Thomas, E. L. Templated Self-Assembly of Block Copolymers: Top-down Helps Bottom-Up. *Adv. Mater.* **2006**, *18* (19), 2505–2521. https://doi.org/10.1002/adma.200502651.

(10) Albert, J. N. L.; Epps, T. H. Self-Assembly of Block Copolymer Thin Films. *Mater. Today* **2010**, *13* (6), 24–33. https://doi.org/10.1016/S1369-7021(10)70106-1.

(11) Ok, S.; Vayer, M.; Sinturel, C. As Featured in :A Decade of Innovation and Progress in Understanding the Morphology and Structure of Heterogeneous Polymers in Rigid Confinement. *Soft Matter* **2021**, *17*, 7430. https://doi.org/10.1039/D1SM00522G.

(12) Wu, Y.; Cheng, G.; Katsov, K.; Sides, S. W.; Wang, J.; Tang, J.; Fredrickson, G. H.; Moskovits, M.; Stucky, G. D. Composite Mesostructures by Nano-Confinement. *Nat. Mater.* **2004**, *3* (11), 816–822. https://doi.org/10.1038/nmat1230.

(13) Kalra, V.; Mendez, S.; Lee, J. H.; Nguyen, H.; Marquez, M.; Joo, Y. L. Confined Assembly in Coaxially Electrospun Block-Copolymer Fibers. *Adv. Mater.* **2006**, *18* (24), 3299–3303. https://doi.org/10.1002/adma.200601948.

(14) Ma, M.; Krikorian, V.; Yu, J. H.; Thomas, E. L.; Rutledge, G. C. Electrospun Polymer Nanofibers with Internal Periodic Structure Obtained by Microphase Separation of Cylindrically Confined Block Copolymers. *Nano Lett.* **2006**, *6* (12), 2969–2972. https://doi.org/10.1021/nl062311z.

(15) Dobriyal, P.; Xiang, H.; Kazuyuki, M.; Chen, J. T.; Jinnai, H.; Russell, T. P. Cylindrically Confined Diblock Copolymers. *Macromolecules* **2009**, *42* (22), 9082–9088. https://doi.org/10.1021/ma901730a.

(16) Wu, Y.; Livneh, T.; Zhang, Y. X.; Cheng, G.; Wang, J.; Tang, J.; Moskovits, M.; Stucky, G. D. Templated Synthesis of Highly Ordered Mesostructured Nanowires and Nanowire Arrays. *Nano Lett.* **2004**, *4* (12), 2337–2342. https://doi.org/10.1021/nl048653r.

(17) Arsenault, A. C.; Rider, D. A.; Tétreault, N.; Chen, J. I. L.; Coombs, N.; Ozin, G. A.; Manners, I. Block Copolymers under Periodic, Strong Three-Dimensional Confinement. *J. Am. Chem. Soc.* **2005**, *127* (28), 9954–9955. https://doi.org/10.1021/ja052483i.

(18) Rider, D. A.; Chen, J. I. L.; Eloi, J. C.; Arsenault, A. C.; Russell, T. P.; Ozin, G. A.; Manners, I. Controlling the Morphologies of Organometallic Block Copolymers in the 3-Dimensional Spatial Confinement of Colloidal and Inverse Colloidal Crystals. *Macromolecules* **2008**, *41* (6), 2250–2259. https://doi.org/10.1021/ma7020248.

(19) Wyman, I.; Njikang, G.; Liu, G. When Emulsification Meets Self-Assembly: The Role of Emulsification in Directing Block Copolymer Assembly. *Prog. Polym. Sci.* **2011**, *36* (9), 1152–1183. https://doi.org/10.1016/j.progpolymsci.2011.04.005.

(20) Higuchi, T.; Tajima, A.; Motoyoshi, K.; Yabu, H.; Shimomura, M. Frustrated Phases of Block Copolymers in Nanoparticles. *Angew. Chemie - Int. Ed.* **2008**, *47* (42), 8044–8046. https://doi.org/10.1002/anie.200803003.

(21) Li, L.; Matsunaga, K.; Zhu, J.; Higuchi, T.; Yabu, H.; Shimomura, M.; Jinnai, H.; Hayward, R. C.; Russell, T. P. Solvent-Driven Evolution of Block Copolymer Morphology under 3D Confinement. *Macromolecules* **2010**, *43* (18), 7807–7812. https://doi.org/10.1021/ma101529b.

(22) Wang, Q.; Yan, Q.; Nealey, P. F.; De Pablo, J. J. Monte Carlo Simulations of Diblock Copolymer Thin Films Confined between Two Homogeneous Surfaces. *J. Chem. Phys.* **2000**, *112* (1), 450–464. https://doi.org/10.1063/1.480639.

(23) Wang, Q.; Nealey, P. F.; De Pablo, J. J. Monte Carlo Simulations of Asymmetric Diblock Copolymer Thin Films Confined between Two Homogeneous Surfaces. *Macromolecules* **2001**, *34* (10), 3458–3470. https://doi.org/10.1021/ma0018751.

(24) Reffner, J. R. The Influence of Surfaces on Structure Formation: I. Artificial Epitaxy of Metals on Polymers. II. Phase Separation of Block Copolymers and Polymer Blends under Nonplanar

Surface Constraints. *Ph.D. Thesis, Univ. Massachusetts Amherst* **1992**.

(25) Yabu, H.; Higuchi, T.; Shimomura, M. Unique Phase-Separation Structures of Block-Copolymer Nanoparticles. *Adv. Mater.* **2005**, *17* (17), 2062–2065. https://doi.org/10.1002/adma.200500255.

(26) He, X.; Song, M.; Liang, H.; Pan, C. Self-Assembly of the Symmetric Diblock Copolymer in a Confined State : Monte Carlo Simulation. *J. Chem. Phys.* **2001**, *114*, 10510. https://doi.org/10.1063/1.1372189.

(27) Yu, B.; Li, B.; Jin, Q.; Ding, D.; Shi, A. Self-Assembly of Symmetric Diblock Copolymers Confined in Spherical Nanopores. *Macromolecules* **2007**, *40* (25), 9133–9142. https://doi.org/10.1021/ma071624t.

(28) Yu, B.; Sun, P.; Chen, T.; Jin, Q.; Ding, D.; Li, B.; Shi, A.-C. Confinement-Induced Novel Morphologies of Block Copolymers. *Phys. Rev. Lett.* **2006**, *96*, 138306.

(29) Li, W.; Wickham, R. A.; Garbary, R. A. Phase Diagram for a Diblock Copolymer Melt under Cylindrical Confinement. *Macromolecules* **2006**, *39* (2), 806–811. https://doi.org/10.1021/ma052151y.

(30) Wang, Y.; Qin, Y.; Berger, A.; Yau, E.; He, C.; Zhang, L.; Gösele, U.; Knez, M.; Steinhart, M. Nanoscopic Morphologies in Block Copolymer Nanorods as Templates for Atomic-Layer Deposition of Semiconductors. *Adv. Mater.* **2009**, *21* (27), 2763–2766. https://doi.org/10.1002/adma.200900136.

(31) Zhao, F.; Xu, Z.; Li, W. Self-Assembly of Asymmetric Diblock Copolymers under the Spherical Confinement. *Macromolecules* **2021**, *54* (24), 11351–11359. https://doi.org/10.1021/acs.macromol.1c02250.

(32) Lodge, T. P.; Pudil, B.; Hanley, K. J. The Full Phase Behavior for Block Copolymers in Solvents of Varying Selectivity. *Macromolecules* **2002**, *35* (12), 4707–4717. https://doi.org/10.1021/ma0200975.

(33) Thomas, A.; Schierhorn, M.; Wu, Y.; Stucky, G. Assembly of Spherical Micelles in 2D Physical Confinements and Their Replication into Mesoporous Silica Nanorods. *J. Mater. Chem.* **2007**, *17* (43), 4558--4562. https://doi.org/10.1039/b702895d.

(34) Larson, R. G. Self-Assembly of Surfactant Liquid Crystalline Phases by Monte Carlo Simulation. *J. Chem. Phys.* **1989**, *91*, 2479–2488.

(35) Larson, R. G. Monte Carlo Simulation of Microstructural Transitions in Surfactant Systems. *J. Chem. Phys.* **1992**, *96*, 7904–7918.

(36) Carmesin, C.; Kremer, K. The Bond Fluctuation Method: A New Effective Algorithm for the Dynamics of Polymers in All Spatial Dimensions. *Macromolecules* **1988**, *21*, 2819–2823.

(37) Kirkpatrick, S.; Gelatt Jr., C. D.; Vecchi, M. P. Optimization by Simulated Annealing. *Science* **1983**, *220*, 671–680.

(38) Grest, G. S.; Soukoulis, C. M.; Levin, K. Cooling-Rate Dependence for the Spin-Glass Ground-State Energy: Implications for Optimization by Simulated Annealing. *Phys. Rev. Lett.* **1986**, *56*, 1148–1151.

(39) Sun, P.; Yin, Y.; Li, B.; Chen, T.; Jin, Q.; Ding, D.; Shi, A. C. Simulated Annealing Study of Morphological Transitions of Diblock Copolymers in Solution. *J. Chem. Phys.* **2005**, *122*, 204905.

(40) Theodorou, D. N.; Suter, U. W. Shape of Unperturbed Linear Polymers: Polypropylene. *Macromolecules* **1985**, *18* (6), 1206–1214. https://doi.org/10.1021/ma00148a028.

(41) Mattice, W. L.; Suter, U. W. *Conformational Theory of Large Molecules: The Rotational Isomeric State Model in Macromolecular Systems*; Wiley-Interscience, 1994.

(42) Berezkin, A. V; Papadakis, C. M.; Potemkin, I. I. Vertical Domain Orientation in Cylinder-Forming Diblock Copolymer Films upon Solvent Vapor Annealing. *Macromolecules* **2016**, *49* (1), 415–424. https://doi.org/10.1021/acs.macromol.5b01771.

(43) Johnson, N. W. Convex Polyhedra with Regular Faces. *Can. J. Math.* **1966**, *18*, 169–200. https://doi.org/10.4153/cjm-1966-021-8.

(44) Wenninger, M. J. *Polyhedron Models*; Cambridge University Press, 1974. https://doi.org/10.1017/CBO9780511569746.